\pdfoutput=1
\documentclass[aps,prx,showpacs,  amsmath,amssymb,longbibliography,showkeys,12pt]{revtex4-2}%
\usepackage{amsfonts}
\usepackage{amsmath}
\usepackage{amssymb}
\usepackage{graphicx}
\usepackage{epsfig}
\usepackage{exscale}
\usepackage{float}
\usepackage{bm}
\usepackage{suffix}
\usepackage{mathtools}
\usepackage{newunicodechar}
\usepackage{lipsum}
\usepackage{enumitem}
\usepackage{bbding}
\usepackage{tikz}
\usepackage{multirow}%
\usepackage[colorlinks=true,linkcolor=blue]{hyperref}
\usepackage{ifpdf }
\expandafter\ifx\csname package@font\endcsname\relax\else
 \expandafter\expandafter
 \expandafter\usepackage
 \expandafter\expandafter
 \expandafter{\csname package@font\endcsname}
\fi
\providecommand{\U}[1]{\protect\rule{.1in}{.1in}}
\DeclarePairedDelimiterX\MeijerM[3]{\lparen}{\rparen}
{\begin{smallmatrix}#1 \\ #2\end{smallmatrix}\delimsize\vert\,#3}
\newcommand\MeijerG[8][]{  G^{\,#2,#3}_{#4,#5}\MeijerM[#1]{#6}{#7}{#8}}
\WithSuffix
\newcommand\MeijerG*
[7]{
G^{\,#1,#2}_{#3,#4}\MeijerM*{#5}{#6}{#7}}

\begin{document}
\title[ ]{ Universal Large-order asymptotic behavior of  the Strong-coupling and High-Temperature series expansions }
\author{Abouzeid M. Shalaby}
\email{amshalab@qu.edu.qa}
\affiliation{Department of Mathematics, Statistics, and Physics, Qatar University, Al
Tarfa, Doha 2713, Qatar}
\keywords{Universality, $\mathcal{PT}$-symmetry, Hypergeometric Resummation, Strong-Coupling expansion}
\pacs{02.30.Lt,64.70.Tg,11.10.Kk}

\begin{abstract}
 For theories that exhibit second order phase transition, we conjecture that the large-order asymptotic behavior of the strong-coupling ( High-Temperature) series expansion takes the form $\sigma^{n} n^{b}$ where  $b$ is a universal parameter. The associated critical exponent is then given by $b+1$. The series itself can be approximated by the hypergeometric approximants $_{p}F_{p-1}$  which can mimic the same large-order behavior of the given series. Near the tip of the branch cut, the hypergeometric function $_{p}F_{p-1}$ has a power-law behavior from which the critical exponent and critical coupling can be extracted. The conjecture has been tested in this work for  the perturbation series of the ground state energy of the Yang-Lee model as a strong-coupling form of the $\mathcal{PT}$-symmetric $i\phi^3$ theory and the High-Temperature expansion within the Ising model. From the known $b$ parameter for the Yang-Lee model, we obtained the exact critical exponents which reflects the universality of $b$. Very accurate prediction for $b$ has been obtained from the many orders available for the High-Temperature series expansion of the Ising model which in turn predicts accurate critical exponent. Apart from critical exponents,  the hypergeometric approximants for the Yang-Lee model show almost exact predictions for
the ground state energy from low orders of perturbation series as input. 
\end{abstract}
\maketitle
\section{ Introduction}
Quantum filed theory represents one of the most successful tools to study
critical phenomena in physics. The point is that one can have different models
behave similarly near the critical point. In this case we say that these
models are in the same class of university where critical exponents (for instance) are the same for the whole class. The Ising model from magnetism and
the $\phi^{4}$ scalar field theory reflect that belief as both are well known
to lie in the same class of universality. Near the critical point, however,
perturbations (weak-coupling) always fail to give reliable results. The reason behind this is
that the effective coupling blows up and turns the theory highly
non-perturbative for which one has to employ rigorous  non-perturbative
techniques to be able to extract reliable results. From a mathematical point of
view, the weak-coupling expansion   diverges because one expands around a  point which represents an essential singularity of the theory \cite{Dayson,singular,Kleinert2009}. Accordingly, one can expect
that the expansion around another point in the coupling space might lead to a
different behavior of the perturbation series. This is what strong-coupling
expansion in field theory \cite{bend-strong,Strong,Kleinert2009} and the equivalent high-temperature (HT) expansion in statistical
systems \cite{HT1, HT2} are expected to do. A note to be mentioned is that the mathematical structure of the strong coupling expansion in  lattice field theory is equivalent to the  high-temperature (HT) expansion in condensed matter physics \cite{Jug1999,Baker1982,Cooper1982,Yamada2007}. Also, lattice spacing serves  as a suitable regularization of the theory under consideration. Being expanded around a point of no essential singularity, the strong-coupling expansion is known to have a finite radius of convergence \cite{poslat} similar to the $_{p}F_{p-1}$ hypergeometric approximants  while the weak-coupling expansion for the $\phi^{4}$ or $i\phi^{3}$ (for instance) field theories has a zero radius of convergence manifested by the $n!$ growth factor in the large-order asymptotic behavior of the series.

Near second order phase transition,   a physical quantity $Q\left(  T\right)  $, where $T$ is the temperature, 
has a power-law behavior of the from  $Q\left(  T\right)  \propto {}_1F_0\left(\psi;\ ;\frac{T}{T_c}\right) =\left(  1-\frac{T}{T_{c}}\right)  ^{-\psi}$
where $\psi$ is called a critical exponent while $T_{c}$ is the critical
temperature. In fact, all the hypergeometric functions ${}_{k+1}F_{k}\left(a_1,a_2,\dots a_{k+1};b_1,b_2,\dots b_{k} ;\frac{T}{T_c}\right)$ have such power-law behavior near the tip of the branch cut which mimics the critical point. The hypergeometric series  ${}_{k+1}F_{k}\left(a_1,a_2,\dots a_{k+1};b_1,b_2,\dots b_{k} ;\frac{T}{T_c}\right)$  has a finite radius of convergence as its large-order asymptotic behavior looks like $\left(  -\frac{1}{T_{c}}\right)  ^{n}n^{b}$ where \cite{Abo-expon}:
\begin{equation}
   b=\psi-1=  \sum_{i=1}^{k+1} a_i-\sum_{j=1}^{k} b_j   
\end{equation}
So it is clear that the large-order parameter $b$ of a series with finite radius of convergence totally defines the critical
exponent and thus is expected to be universal, the same way  critical exponents do. 

The power-law behavior characterizing the   second order phase transition is  itself   the member ${}_1F_0$ of the set of hypergeometric approximants ${}_{k+1}F_{k}\left(a_1,a_2,\dots a_{k+1};b_1,b_2,\dots b_{k} ;\frac{T}{T_c}\right)$. Accordingly, the critical exponent can be shown easily to be related to the large-order asymptotic behavior of the expansion of the power- law formula. Let
us rewrite the critical behavior in a another but equivalent form:%
\begin{align*}
Q\left(  T\right)    & \propto\left(  T-T_{c}\right)  ^{-\psi}\\
& =T^{-\psi}\left(  1-\frac{T_{c}}{T}\right)  ^{-\psi}.
\end{align*}
For non-trivial transition $\left(  T_{c}\neq0\right)$, we can have a high
temperature expansion of $Q(T)$ which again has a finite-radius of convergence with the
parameter $b$ again is given by $b=\psi-1$.  The relation $b=\psi-1$, although proved by considering hypergeometric approximants,   is in fact general and is a manifestation of the theorem  of Darboux which implies that late terms of the expansion of a power-law form and that of a Taylor series of the given quantity  are of the same form \cite{darboux} provided that the series has a finite radius  of convergence.

Our conjecture for the existence of an expansion with a universal $b$ parameter motivates for the study of second-order phase transition within the strong-coupling expansion in quantum field theory. The fact that the strong-coupling (HT) expansion has a finite radius of convergence will be stressed in sec.\ref{large-strong}.  Our idea for the preference to study critical phenomena within strong-coupling expansion  is very important as there are quasi-classical techniques that are supposed to obtain the exact large-order asymptotic behavior of that expansion.  Thus, for the sake of getting the first exact critical exponent in three dimensions, it is   worth it to make   the needed effort to study the strong-coupling expansion in field theory. Such kind of studies   can relate critical exponents to the asymptotic large-order parameter $b$ which we  expect  to be universal.  

 In this paper,   we shall stress the Yang-Lee model as a strong-coupling expansion of a field theory in $0+1$ dimensions. For that model,  the asymptotic large-order behavior  is known and and thus can be used to show that the parameter $b$ is universal. Taking into account that  the  HT-expansion in condensed matter and strong-coupling from lattice field theory are two sides of the same coin \cite{Yamada2007,Cooper1982,Butera97,Baker1982, bend-d,HT-STM,arisue87,HT-ST-Sav5} and that the high-temperature (strong-coupling) expansion is known up to a relatively high order, we shall stress that expansion for   both SQ and SC lattices for  the  Ising model and show again that the parameter $b$ is universal. 

The rest of this paper is organized as follows. In sec.\ref{large-strong}, we highlight the fact that the strong-coupling (HT) expansion has a finite radius of divergence. In sec \ref{parametrization}, the weak-coupling, strong-coupling and large-order parametrization of the hypergeometric approximants is stressed.  In sec.\ref{Yangs}, we apply the hypergeometric approximation for the series of   the ground state-energy of the Yang-Lee model. In this section, all critical exponents are obtained exactly from knowing the $b$ parameter in the large-order behavior.  In sec.\ref{HT-SC}, the HT-expansion of the susceptibility  within the SC lattice of the Ising-model is investigated while the SQ case is investigated in sec.\ref{HT-SQ}. Using  the   last highest orders (large $n$) of the known $25$ orders of the associated perturbation series, we were able to obtain very accurate approximation for the parameter $b$ which in turn shows its universality via comparison with the well known results for the  $\gamma$ exponent.  Summary and conclusions will follow in sec.\ref{conc}. 

\section{Large-order asymptotic behavior from  the strong-coupling expansion in quantum field theory}\label{large-strong}

 There is a one-to-one correspondence between the $n!$ growth factor in the large-order asymptotic behavior of the week-coupling
expansion and the essential singularity existing at zero coupling \cite{Kleinert2009,singular}. For the $\phi^{4}$ scalar field theory, for
instance, the large-order asymptotic behavior for the weak-coupling expansion
takes the form $n!\sigma^{n}n^{b}$. In Refs.\cite{Abo-expon,Abo-large,Abo-expon7}, 
we showed that a series of such behavior ( it has a zero-radius of convergence) can be fitted by the hypergeometric
approximants $_{\text{ }p+1}F_{p-1}(a_{1},...a_{p+1};b_{1}....b_{p-1};\sigma z)$. These hypergeometric approximants    can be analytically continued to non-zero $z$ values via their
representation in terms of the Meijer G function. On the other hand, the strong-coupling
(High-Temperature) expansion   is well known to have a finite
radius of convergence \cite{Kleinert2009,poslat} and thus  the asymptotic large-order behavior is taking the form
$\sigma^{n}n^{b}$ without an $n!$ growth factor found in the week-coupling expansion. In Ref.\cite{Abo-expon}, we
showed that such type of series ( with finite radius of convergence) can be approximated by  a different type of hypergeometric approximants ( $_{\text{ }p+1}F_{p}(a_{1},...a_{p+1};b_{1}....b_{p};\sigma z)$). These approximants can
produce the same large-order asymptotic behavior with their parameters are constrained
as :%
\begin{equation}
b+1=\sum_{i=1}^{p+1}a_{i}-\sum_{j=1}^{p}b_{j}.
\end{equation}
Near the branch cut, the approximants  $_{\text{ }p+1}%
F_{p}(a_{1},...a_{p+1};b_{1}....b_{p};\sigma z)$ have a power-law behavior of
the form:
	\[ _{\text{ }p+1}F_{p}(a_{1},...a_{p+1};b_{1}....b_{p};\sigma z)\sim\left(
1-\sigma z\right)^{-\psi}, 
\]
where $\psi=b+1$. Accordingly, knowing the large order parameter $b$ of the
strong-coupling expansion will lead to the exact determination of the critical
exponent while knowing the parameter $\sigma$ will determine the critical coupling.

In literature, there exist quasi-classical techniques (out of the scope of this work) for the exact
determination of the large-order parameters $b$  and $\sigma$
\cite{Kleinert-Borel,Kleinert2009}. However, up to the best of our knowledge
this issue has not been stressed rigorously for the strong-coupling expansion in field theory in other than one dimensional cases (quantum mechanics).
Fortunately, strong-coupling (High-Temperature) expansions for many models are
listed in literature up to high orders \cite{STL01,xyht1,HT2,HT1,Butera97} which means that one can extract approximate values of the parameters $b$ and
$\sigma$. So at least  approximately,  one can  test the validity of our conjecture about the universality of $b$ by extracting this parameter from the relatively high number of terms available in literature. Confirming the universality of $b$ might open the door for the first determination of exact critical exponents from the exact determination of $b$ for the strong-coupling (HT) expansion.

 Before we try to test our conjecture, let us first highlight the fact that the strong-coupling (High-temperature) expansion possesses a finite radius of convergence and thus at large $n$, the the $n^{th}$ coefficient behaves like $ \sigma^n n^b$.  To do that, in the following,  we list different Hamiltonian models for which the strong coupling expansions can be shown to have a finite radius of convergence:

\subsection{Large-order asymptotic behavior of the strong-coupling expansion for anharmonic oscillators}

The  Hamiltonian of the anharmonic oscillators is given by
\begin{equation}
H_{s}=   p^{2}+x^{2}+\beta x^{2m}.
\end{equation}
Here $\beta$ is the coupling constant (should not be confused with inverse temperature $\beta$). The strong-coupling of this Hamiltonian has been stressed in Ref.\cite{Large-strong} and the ground state energy has been shown to have an
expansion of the form:

\begin{equation}
E_{0}=\beta^{\frac{1}{m+1}}\sum_{n=0}^{\infty}h_{n}\beta^{\frac{-2n}{m+1}},
\end{equation}
where for the limit $\text{ \ }n\rightarrow\infty$ we have the asymptotic
form:
\begin{equation}
h_{n}\sim c\ n^{-\frac{3}{2}}\sigma^{n}\left(  1+O\left(  \frac{1}{n}\right)
\right)  .\\ \nonumber
\end{equation}
The ratio test can tell us clearly that the strong-coupling series above has a
finite radius of convergence with the parameter $b$ having its exact value of  $-3/2$.  One can realize that   $b=\frac{-3}{2}$ for
the different Hamiltonians (different $m$ values) in the set. It is well known that at the Ising limit ($\beta\rightarrow\infty$), physical quantities behave similarly for different interaction Hamiltonian \cite{bend-d}. As long as $b$ is same for different $m$, one might conclude that it is a universal quantity.   Also one can
conclude that $b=\frac{-3}{2}$ for the $\mathcal{PT}$-symmetric $ix^{3}$
Hamiltonian as well. In the next section we shall see that this value determines the known exact critical exponent for that model.

 Since $b$ defines the associated critical exponent and so far is not known exactly for field theories in higher dimensions, we shall
try to obtain an approximation for $b$ from the strong-coupling expansion of the $\mathcal{PT}$-symmetric $ix^{3}$ model and compare it with the exact result. After that we can extend the same strategy for the strong-coupling expansion in field theory to get an approximate value for $b$ and thus test its universality.

\begin{figure}[htp]
\begin{center}
\epsfig{file=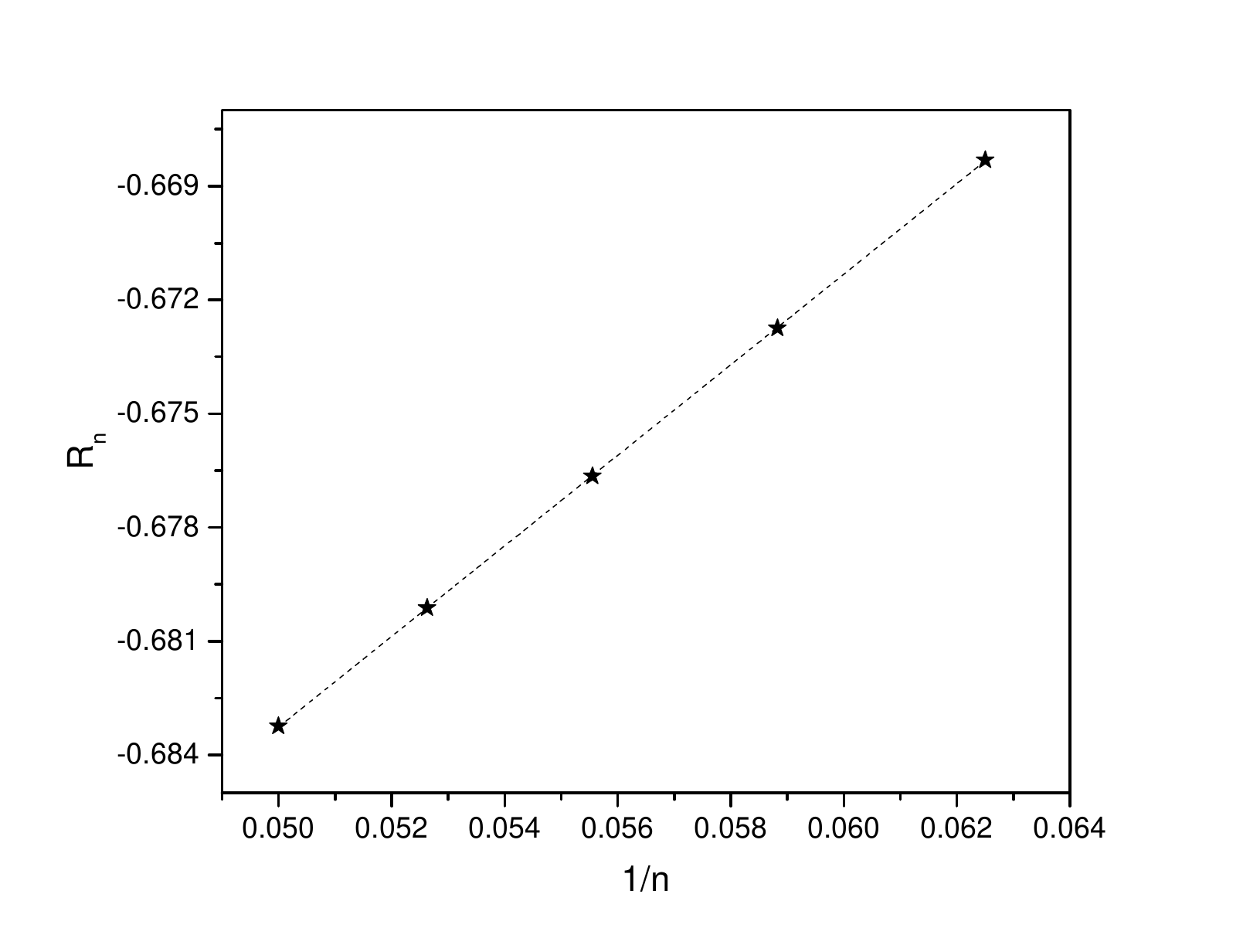,width=0.65\textwidth}
\end{center}
\caption{\textit{In this figure, we plot the coefficients ratio $R_n$ for the  strong-coupling expansion of the vacuum energy of the $\mathcal{PT}$-symmetric $ix^{3}$ model (obtained in Ref.\cite{zin-borel}) at large orders. The data (stars) has a straight line fit (dashed) of the form $R_n=1.1937(\frac{1}{n})-0.7429$ which predicts the values $\sigma=-0.7429$ and $b=-1.6068$.}} 
\label{x3-efit} 
\end{figure}

 For   large $n$, the ratio $R_n=h_n/h_{n-1}$ can be approximated as :
	 \begin{align*}
R_{n}  & =\frac{cn^{b}\sigma^{n}}{c\left(  n-1\right)  ^{b}\sigma^{n-1}%
}=\sigma\left(  1-\frac{1}{n}\right)  ^{b}\\
& \simeq\sigma-b\sigma(\frac{1}{n}).
\end{align*}
Accordingly, $R_{n}$ for  series with finite radius of convergence can be
fitted with a straight line when plotted versus $\frac{1}{n}$ (for  large
$n)$. The first twenty coefficients of the series in Eq.(\ref{Jpert}) in Sec.\ref{Yangs} have been obtained in Ref.\cite{zin-borel} ( Eq.(92) there). In
Fig.\ref{x3-efit},  we plotted $R_{n}$ versus $\frac{1}{n}$ where the data can be fitted by the equation:
\[
R_{n}=1.1937(\frac{1}{n})-0.7429.
\]
From this equation   one can conclude approximate  values for the large order parameters
as $b=-1.6068$ compared to its exact value $b=\frac{-3}{2}$ and
$\sigma=0.7429$. In sec.\ref{Yangs}, we shall see that fitting the series using
the hypergeometric approximants $_{\text{ }p+1}F_{p}(a_{1},...a_{p+1}%
;b_{1}....b_{p};\sigma z)$ can give better values for $\sigma$ which
determines the critical coupling.
The plot in Fig.\ref{x3-efit} is thus showing that one can get approximate values for the parameters $b$ and $\sigma$ from the last few orders in the strong-coupling expansion.
\subsection{ Strong-coupling expansion for the $g \phi_{1+1}^{4}$ vacuum energy}

\begin{figure}[htp]
\begin{center}
\epsfig{file=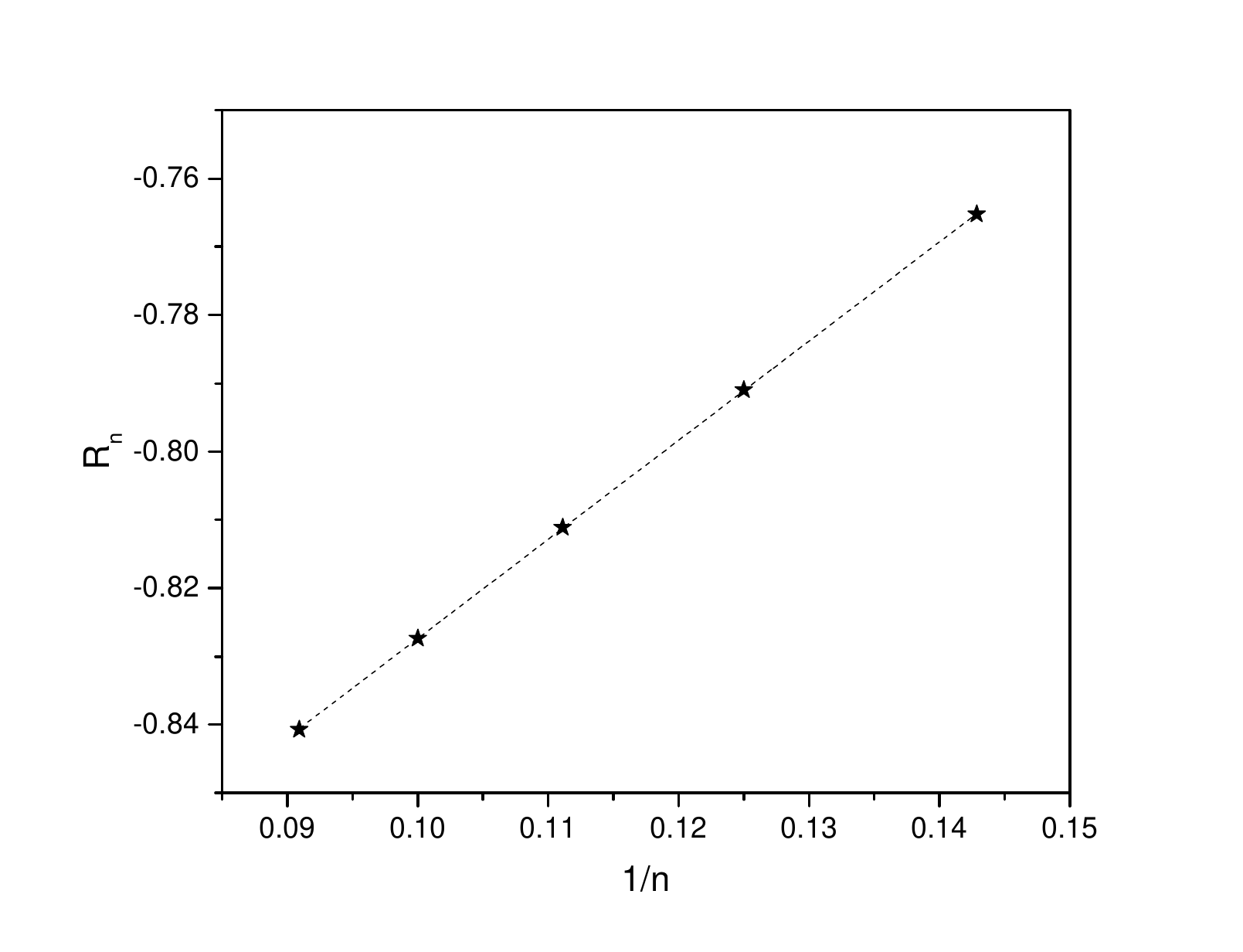,width=0.65\textwidth}
\end{center}
\caption{\textit{In this figure, we plot the ratio $R_n$ for the  strong-coupling expansion (lattice)  of the vacuum energy for the ${\phi^4}_{1+1}$ scalar field theory at large n. The data has a straight line fit of the form $R_n=1.4524(\frac{1}{n})- 0.97261$ which predicts the values $\sigma=0.9726$ and $b=-1.4933$.}} 
\label{Phi41-1} 
\end{figure}

In Ref.\cite{STL01}, strong-coupling series (lattice) for the vacuum energy of
the $g\phi_{1+1}^{4}$scalar filed theory has been obtained up to $11^{th}$ order
in $y=g^{-\frac{2}{3}}$ as:%
\begin{align*}
E  & =g^{\frac{1}{3}}e_{g},\\
e_{g}  & \simeq
0.66798625915577710827096201688+0.43100635014259473006095738275\lambda\\
& +{\dots\dots\dots\dots\dots}-0.0087493465269972\lambda^{8}+0.007096747591805\lambda
^{9}-0.005871428\lambda^{10}\\
& +0.0049362\lambda^{11}.
\end{align*}
The ratio test can confirm the convergence of the series for $e_{g}$. We plot
$R_{n}$ versus $\frac{1}{n}$ in Fig.\ref{Phi41-1} where the fitting gives  $\sigma=0.9726$ and
$b=-1.4933$. These results can be checked by the radius of
convergence $\frac{1}{\sigma}= 1.0282$ which is
very close to the findings in Ref.\cite{STL01}. Of course the hypergeometric
approximants $_{\text{ }p+1}F_{p}(a_{1},...a_{p+1};b_{1}....b_{p}%
;\sigma\lambda)$ can be parametrized to fit the given series with better
approximation for the values of $\sigma$ and $b$.

\subsection{High-Temperature (Strong-coupling) expansion for the $ O(2)$-symmetric $\phi_{2+1}^{4}$ model}

\begin{figure}[htp]
\begin{center}
\epsfig{file=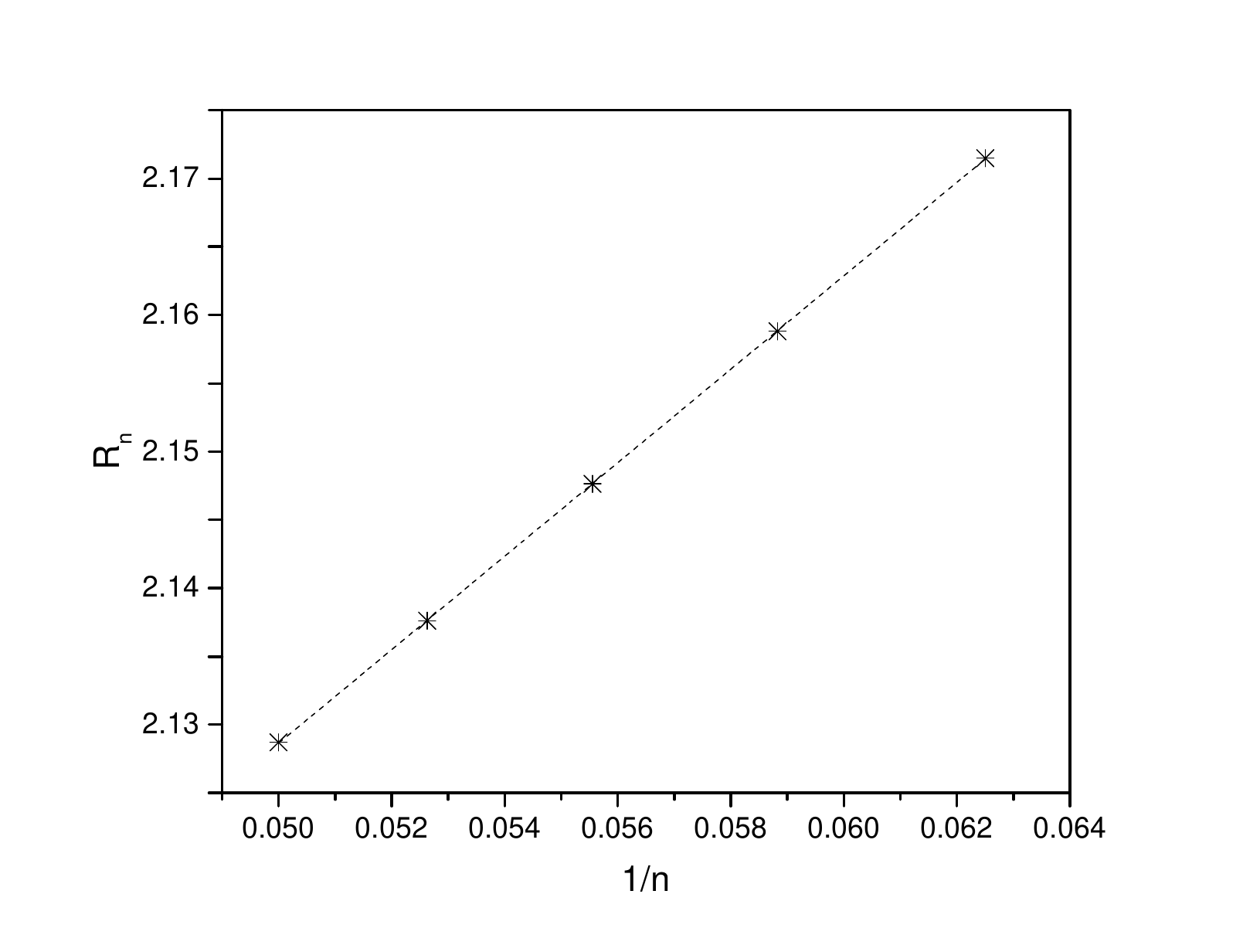,width=0.65\textwidth}
\end{center}
\caption{\textit{The plot of  $R_n$ for the  High-Temperature (strong-coupling) expansion (lattice)  of the second moment correlation function of the
of the three-dimensional $\phi^4$ field theory with $O(2)$ symmetry. The data can be fitted as  $R_n=3.4227(\frac{1}{n})+1.9575$ which predicts the values $\sigma=1.9575$ and $b=-3.4227$.}} 
\label{xy-m2-3d-phi} 
\end{figure}

In Ref.\cite{xyht1}, the High-Temperature (strong-coupling) expansion for the
second moment of the two-point function for the $O(2)$-symmetric  $\phi_{2+1}^{4}$ scalar
field theory is listed up to $\beta^{20}$, where $\beta$ is the inverse
temperature (column 2 in TABLE  XVII  there). In Fig.\ref{xy-m2-3d-phi}, we generated the plot for $R_n$ versus $1/n$
and extracted the values $b=1.7485$ while $\sigma=1.9575$. The
critical inverse temperature $\beta_{c}=\frac{1}{\sigma}=0.51086$ compared to the
result $\beta_{c}=0.5099049$ in Ref.\cite{xyht1}. Again, the hypergeometric approximants are
expected to give better predictions but we will not stress it here as our aim
from this section is to highlight the fact known from literature that the
strong-coupling expansion has a finite radius of convergence and thus having
a large order asymptotic behavior similar to that of  the hypergeometric series $_{\text{ }%
p+1}F_{p}(a_{1},...a_{p+1};b_{1}....b_{p};\sigma \beta)$.  

\section{Weak coupling, Strong-coupling and Large-order parametrization of the
hypergeometric resummation}\label{parametrization}

 Famous non-perturbative tools that are always used in literature to study
critical phenomena in physics are Borel, Borel-Pad$\acute{e}$ and Borel with
conformal mappings resummation algorithms
\cite{zinjustin,Kleinert-Borel,ON17,zin-exp,x3-4l,zin-cr,Eta4,Guillou}. In
applying these algorithms one may face slow convergence and most of the
calculations are achieved using numerical steps. Recently, the simple but
accurate hypergeometric resummation algorithm has been introduced which is of
closed form \cite{Prl}. In Ref.\cite{Abo-hyper}, we showed that one can employ
the strong-coupling data to determine all the numerator parameters in the
hypergeometric $_{p}F_{p-1}$ approximants. In another work
\cite{Abo-large,Abo-expon}, we showed that the relation between $p$ (number of
numerator parameters) and $q$ (number of denominator parameters) is
constrained by the large-order behavior of the given perturbation series. Once
we determined the difference $p-q$ from the large order behavior, one can
employ the large order parameters to accelerate the convergence of the
hypergeometric resummation.

Based on the large order-behavior of a given perturbation series, one can
categorize the divergent series into different classes, where each class can be
resummed by a Hypergeometric function with expansion that reflects the needed
growth factor \cite{Abo-large,Abo-expon}. To clarify this point more, consider
a divergent series for a physical quantity $Q(z)=\sum_{n=0}^{n}c_{n}z^{n}$
that has the following large-order behavior:%
\begin{equation}
c_{n}\sim\alpha\left(  \left(  p-q-1\right)  n\right)  !(-\sigma)^{n}%
n^{b}\left(  1+O\left(  \frac{1}{n}\right)  \right)  ,\text{ \ \ }%
n\rightarrow\infty. \label{LOB}%
\end{equation}
Such types of asymptotic large-order behavior suggest the following approximants:

\begin{enumerate}
\item For a series with finite radius of convergence, we have $\left(
p-q-1\right)  =0$. The suitable hypergeometric approximant is \ then%
\begin{align*}
Q(z)  &  \approx c_{0\text{ }p}F_{p-1}(a_{1},...a_{p};b_{1}....b_{p-1};-\sigma
z), \text{where}\\
b\bigskip &  =\sum_{i=1}^{p}a_{i}-\sum_{i=1}^{p-1}b_{i}-1.
\end{align*}

\item For a divergent series with zero radius of convergence and $n!$ growth
factor ($p-q-1=1$), the suitable approximant is then%
\begin{align*}
Q(z)  &  \approx c_{0\text{ }p}F_{p-2}(a_{1},...a_{p};b_{1}....b_{p-2};-\sigma
z),\\
b\bigskip &  =\sum_{i=1}^{p}a_{i}-\sum_{i=1}^{p-2}b_{i}-2.
\end{align*}

\item For a divergent series with zero-radius of convergence but $\left(
2n\right)  !$ growth factor ($p-q-1=2$) ( the ground state energy of the sixtic
oscillator for instance) then the suitable approximant is
\begin{align*}
Q(z)  &  \approx c_{0\text{ }p}F_{p-3}(a_{1},...a_{p};b_{1}....b_{p-3};-\sigma
z)
\end{align*}
and so on. Of course for $p\geq q+2$, the series $_{\text{ }p}F_{q}$ is
divergent and has a zero-radius of convergence but analytic continuation to
non-zero $z$ values can be offered by a Mellin-Barnes integral representation
of $_{\text{ }p}F_{q}$ or equivalently in terms of the Meijer-G function
\cite{Abo-expon, Abo-large}.

We call the above algorithm the Hypergeometric- Meijer resummation
\cite{Abo-large,Abo-expon}. Note that, in this algorithm once you select the
suitable Hypergeometric approximant based on the growth factor in the large
order behavior, it can accommodate all weak-coupling, strong coupling and
large-order data associated with the given perturbation series. In fact, in
Ref.\cite{Prd-GF}, Mera et.al used Borel-hypergeometric algorithm with Borel
functions of the form $c_{0\text{ }p}F_{p-1}(a_{1},...a_{p};b_{1}%
....b_{p-1};.\sigma z)$. That algorithm results in a Meijer-G function
resummation approximant that employs low order perturbation data as input. In
our technique, we do not use any Borel or Pad$\acute{e}$ methods but instead
we start from large order behavior and select the appropriate hypergeometric
approximant. In case $p-q$ is greater than one, we use the Meijer-G function
representation of the Hypergeometric function \cite{HTF} where%

\begin{equation}
_{\text{ }p}F_{q}(a_{1},...a_{p};b_{1}....b_{q};z)=\frac{\prod_{k=1}^{q}%
\Gamma\left(  b_{k}\right)  }{\prod_{k=1}^{p}\Gamma\left(  a_{k}\right)  }%
\MeijerG*{1}{p}{p}{q+1}{1-a_{1},\dots,1-a_{p}}{0,1-b_{1},\dots,1-b_{q}}{z}.
\label{hyp-G-C}%
\end{equation}

\end{enumerate}
The Meijer G function on the right hand side of this equation has the integral
representation of the form:\cite{HTF}:
\begin{equation}
\MeijerG*{m}{n}{p}{q}{c_{1},\dots,c_{p}}{d_{1},\dots,d_{q}}{z}=\frac{1}{2\pi i}%
\int_{C}\frac{\prod_{k=1}^{n}\Gamma\left(  s-c_{k}+1\right)  \prod_{k=1}%
^{m}\Gamma\left(  d_{k}-s\right)  }{\prod_{k=n+1}^{p}\Gamma\left(
-s+c_{k}\right)  \prod_{k=m+1}^{q}\Gamma\left(  s-d_{k}+1\right)  }z^{s}ds.
\label{hyp-G-C2}%
\end{equation}
By selecting the contour $C$ to run from from $-i\infty$ to $+i\infty$
\cite{HTF}, the integral above converges for $p+q<2(m+n)$. It is then clear that for $p=q+1$ where the hypergeometric series has a finite radius of convergence, the  condition for the convergence of the above integral is satisfied.

The algorithm has been shown to give accurate results for different divergent
series like the ground state energy of anharmonic oscillator
\cite{Abo-hyper,Abo-large} and the critical exponents of the $O(N)$-symmetric
model \cite{Abo-expon,Abo-expon7}. In this work, however, we will concentrate on resummation of strong-coupling (High-Temperature ) series expansion   for systems that show up second order phase transition. Such type of series     have a finite radius of convergence and thus hypergeometric approximants ${}_pF_q$ with $p-q-1=0$ are relevant ones. Near the tip of the branch cut, the approximants posses a power-law behavior where the critical exponents are solely determined by the large-order parameter $b$ while  $\sigma$ determines the critical coupling (or temperature).
\section{Hypergeometric Approximation for the Strong-coupling expansion of the
Yang-Lee model}\label{Yangs}

In 1952, Lee and Yang introduced a theory of phase transitions that is based
on the zeros of the partition function in the complex plane of an external
parameter like the external magnetic field \cite{Young-Lee1,Young-Lee2}. At
the continuum limit, the zeros of the partition function can touch the real
axis which then represents a critical point called Yang-Lee edge
singularity. For many years the zeros of partition function is considered as a
theoretical issue but recently it has been exposed to experimental
investigations (see Ref.\cite{yang-lee-exp} and references therein). In fact
the zeros of the partition function are always existing for non-real external
parameters and thus the theory near the zeros can turn to be non-Hermitian but
$\mathcal{PT}$-symmetric \cite{Abo-novel}. The link between critical behavior
of the Ising model near edge singularity and $\mathcal{PT}$-symmetric
$i\phi^{3}$ theory was first introduced by Fisher who identified an effective
action of the magnetization of the Ising model as a Landau-Ginzberg theory
given by a $\mathcal{PT}$-symmetric $i\phi^{3}$ theory \cite{LYsing1}. We will
study this model in $0+1$ space-time dimension and tackle the critical
behavior associated to the edge singularity from the point of view of the
dependance of the order parameter on the external magnetic field rather than
investigating the zeros of partition function.

Near the edge singularity, perturbative calculation within the Yang-Lee
quantum field model can't account for the expected phase transition. In this
model the Lagrangian density is given by:
\begin{equation}
\mathcal{L}\left[  \phi\right]  =\frac{1}{2}\left(  \partial\phi\right)
^{2}-\frac{1}{2}m^{2}\phi^{2}(x)-\frac{ig}{3}\phi^{3}\left(  x\right)
+iJ\phi\left(  x\right)  .
\end{equation}
We studied this model in Ref.\cite{abo-yang} and showed that in $d=6-\epsilon$
dimensions, there exists a Gaussian fixed point where exact critical
exponents are extracted from the one-loop effective potential. In the same
reference we showed that for dimensions $d<6$, the effective coupling
($\frac{g}{M^{3-\frac{1}{2}d}}$) blows up and the the theory has
non-perturbative fixed point. In these cases, the one-loop effective potential
would not be able to produce reliable results near the critical point. The
worst case exists for $d=1$, where at the critical point ($M\rightarrow0$) the
effective coupling blows up very fast. We used the effective potential to
study this case but faraway from the critical region in Ref.\cite{abo-eff}.
This theory is $\mathcal{PT}$-symmetric \cite{bendr,bendrII} and the $\mathcal{PT}%
$-symmetry is broken at the fixed point \cite{Bend-cr1,Bend-cr2}. At this
point there exists a phase transition at which we showed ( for $d=6-\epsilon$)
\cite{abo-yang} that the fixed point is really representing a Yang-Lee edge
singularity
\cite{ncj,npj,pra,pra1,Opt,LYsing1,Musardob,Young-Lee1,Young-Lee2,Cardy,Cardy2,Wipf}.

The effective action of the magnetization of the Ising model has a
Landau-Ginzburg representation at the continuum limit of the form
\cite{LYsing1}%
\begin{equation}
S=\int dx^{d}\left(  \frac{1}{2}\left(  \partial_{\mu}\phi\right)
^{2}+i\left(  h-h_{c}\right)  \phi+ig\phi^{3}\right)  , \label{Lee-Yang-M}%
\end{equation}
which is equivalent to  the Yang-Lee model above. The critical exponents
associated with the Yang-Lee edge singularity have been listed in
Ref.\cite{LYsing1}. The study of that reference relied on considering the
density of zeros of the partition function which has been shown to follow a
power-law behavior near the edge singularity exactly the same manner the
magnetization follows with respect to the external magnetic field.

The $\mathcal{PT}$-symmetric Yang-Lee model in $0+1$ space-time dimension (quantum mechanics) has
been studied in Ref.\cite{zin-borel}. The Hamiltonian of that model in one
dimension is given by:
\begin{equation}
H_{g}=\frac{\pi^{2}}{2}\  +\frac{1}{2}m^{2}\phi^{2}+\frac{i\sqrt{g}}{6}\phi^{3}.
\label{gx3}%
\end{equation}
The weak-coupling series expansion of the  ground state energy of that model is divergent and thus resummation
techniques are to be followed to get reliable results \cite{zin-borel}. \ A
strong coupling representation can be obtained using a scale and shift
transformations \cite{zin-borel,Abo-hyper} that leads to the form:%
\begin{equation}
H_{g}=\sqrt[5]{g}\left(  \frac{\pi^{2}}{2}+\frac{i\phi^{3}}{6}+\frac{1}%
{2}\frac{im^{4}}{g^{\frac{4}{5}}}\phi\right)  -\frac{m^{6}}{3g}. \label{Jx3}%
\end{equation}
This Hamiltonian can be rewritten as $H_{g\text{ }}=g^{\frac{1}{5}}   H_{J} -\frac{m^{6}}{3g}  $ where
\begin{equation}
H_{J\text{ }}=\frac{\pi^{2}}{2}+\frac{i\phi^{3}}{6}+\frac{1}{2}iJ\phi,
\label{Hamilt}%
\end{equation}
with $J=\frac{im^{4}}{g^{\frac{4}{5}}}$. The Hamiltonian $H_{J}$ has been
studied also in Ref. \cite{zin-borel} where the ODM method is used to resum
the divergent series representing the ground state energy $E_{0}^{J}$ where
\begin{align}
E_{0}^{J}  &  =\sum_{n=0}^{\infty}d_{n}J^{n}%
=.3725457904522070982506011+0.3675358055441936035304J\label{Jpert}\\
&  +0.1437877004150665158339J^{2}+O\left(  J^{3}\right)  .\nonumber
\end{align}
As we explained in the introduction, near critical point physical quantities follow a power-law behavior. The power law behavior of the form $(J-J_c)^\delta$ has a series expansion with finite radius of convergence and a large-order asymptotic behavior like $ \sigma^n n^b$, where $b=-\delta-1$ and $J_c=1/\sigma$. According to the theorem of Darboux, the large order terms in an expansion originates from the singularity (closest to origin) of the expanded function \cite{darboux}. Accordingly, one can expect that   large-order form $\sigma^n n^b$ of the singular part (power-law) and the large-order asymptotic behavior of the whole series (assumed to have a finite radius of convergence like the power-law form)   have the same form.  Accordingly, one concludes the direct relation between the parameter $b$ and the critical exponent. Note that the  weak-coupling expansion has an essential singularity and thus Darboux theorem is not applicable for that case.

The most suitable hypergeometric approximant for a perturbation series is determined
from the large order behavior of that series. In fact, for a class of
interaction Hamiltonian $\beta x^{m}$, the large order behavior has been
obtained in Ref.\cite{Large-strong}. In fact, the large order behavior for the
Hamiltonian $H_{J\text{ }}$ can be concluded from that reference if we set
$m=\frac{3}{2}$ in the Hamiltonian there where:
\[
H_{m}=p^{2}+x^{2}+\beta x^{2m}.
\]
The ground state energy of the rescaled Hamiltonian $\beta^{\frac{2}{5}}H_{m}$
has the expansion%

\[
E_{0}^{m}=\sum_{n=0}^{\infty}h_{n}\beta^{\frac{-2n}{m+1}},
\]
where for the limit $\text{ \ }n\rightarrow\infty$ we have the asymptotic form:
\begin{equation}
h_{n}\sim cn^{-\frac{3}{2}}\sigma^{n}\left(  1+O\left(  \frac{1}{n}\right)
\right). \label{large}%
\end{equation}
Note that the parameter $b=-\frac{3}{2}$ in the large order above does not
depend on $m$ which reflects a kind of universality of the whole class and
thus we can extend it to the case of the Yang-Lee model represented by the
perturbation series in Eq.(\ref{Jpert}). As we expected, this large order behavior tells us that the strong-coupling series expansion has a finite radius of convergence. What is important in the above large-order behavior is that the hypergeometric function $_{\text{ }p}%
F_{p-1}(a_{1},...a_{p};b_{1}....b_{p-1};-\sigma z)$ has the same form of large
order behavior of its expansion. This can be shown by noting that:
\[
_{\text{ }p}F_{p-1}\left(  {a_{1},......a_{p};b_{1},........b_{q};-\sigma
z}\right)  =\sum_{n=0}^{\infty}\frac{\frac{\Gamma\left(  a_{1}+n\right)
}{\Gamma\left(  a_{1}\right)  }....\frac{\Gamma\left(  a_{p}+n\right)
}{\Gamma\left(  a_{p}\right)  }}{n!\frac{\Gamma\left(  b_{1}+n\right)
}{\Gamma\left(  b_{1}\right)  }....\frac{\Gamma\left(  b_{p-1}+n\right)
}{\Gamma\left(  b_{p-1}\right)  }}\left(  -\sigma z\right)  ^{n},
\]
and thus has a large order behavior of the form in Eq.(\ref{large}) but with%
\begin{equation}
\frac{\frac{\Gamma\left(  a_{1}+n\right)  }{\Gamma\left(  a_{1}\right)
}....\frac{\Gamma\left(  a_{p}+n\right)  }{\Gamma\left(  a_{p}\right)  }%
}{n!\frac{\Gamma\left(  b_{1}+n\right)  }{\Gamma\left(  b_{1}\right)
}....\frac{\Gamma\left(  b_{p-1}+n\right)  }{\Gamma\left(  b_{p-1}\right)  }%
}\left(  -\sigma\right)  ^{n}\sim\gamma\ (-\sigma)^{n}n^{b}\left(  1+O\left(
\frac{1}{n}\right)  \right)  ,\text{ \ \ }n\rightarrow\infty,
\end{equation}
where
\begin{equation}
\sum_{i=1}^{p}a_{i}-\sum_{i=1}^{p-1}b_{i}-1=b,
\end{equation}
and
\[
\gamma=\frac{%
%TCIMACRO{\dprod \limits_{i=1}^{p-1}}%
%BeginExpansion
{\displaystyle\prod\limits_{i=1}^{p-1}}
%EndExpansion
\Gamma\left(  b_{i}\right)  }{%
%TCIMACRO{\dprod \limits_{i=1}^{p}}%
%BeginExpansion
{\displaystyle\prod\limits_{i=1}^{p}}
%EndExpansion
\Gamma\left(  a_{i}\right)  }.
\]
We can obtain the above relations easily using the asymptotic form of a ratio
of two $\Gamma$ functions \cite{Gamma}:%

\begin{equation}
\frac{\Gamma\left(  n+\alpha\right)  }{\Gamma\left(  n+\beta\right)
}=n^{\alpha-\beta}\left(  1+\frac{\left(  \alpha-\beta\right)  \left(
-1+\alpha+\beta\right)  }{n}+O\left(  \frac{1}{n^{2}}\right)  \right)  .
\end{equation}
 Since the hypergeometric function $_{\text{ }p}F_{p-1}\left(  {a_{1},......a_{p};b_{1},........b_{p-1};\sigma z}\right)  $ can reproduce the same
form of large-order behavior of the perturbation series under consideration,
it is then recommended as an approximant for the perturbation series of
$E_{0}^{J}$ above. Near the tip of the branch cut, the hypergeometric function has a power-law behavior of the form:

\begin{equation}
 _{\text{ }p}%
F_{p-1}\left(  {a_{1},......a_{p};b_{1},........b_{p-1};\sigma z}\right)\propto (1-\sigma z)^{-\psi},
\end{equation}
where $\psi=\sum_{i=1}^{p}a_{i}-\sum_{i=1}^{p-1}b_{i}=b+1$ or in other words, the critical exponent $\psi$ is solely determined by the large-order parameter $b$ which means that this parameter is universal.

 Based on the above clarifications,  the hypergeometric resummation algorithm can be simplified into  two simple steps:

\begin{enumerate}
\item Match the available orders from the perturbation series with the
corresponding number of terms from the expansion of $_{\text{ }p}%
F_{p-1}\left(  {a_{1},......a_{p};b_{1},........b_{p-1};\sigma J}\right)  .$

\item Employ the large order relation
\begin{equation}
\sum_{i=1}^{p}a_{i}-\sum_{i=1}^{p-1}b_{i}-1=-\frac{3}{2}, \label{large-b}%
\end{equation}
in the set of coupled equations to obtain the $b_{i}$ parameters. Note that
the $a_{i}$ parameters for the model under consideration are known
\cite{zin-borel}.
\end{enumerate}
Let us give an example for a certain order of the hypergeometric approximant.
Assume that we have the second order perturbation series of the form:%

\begin{equation}
Q\left(  z\right)  =c_{0}+c_{1}z+c_{2}z^{2}+O\left(  z^{3}\right)  ,
\label{Qpert}%
\end{equation}
with the large order behavior in Eq.(\ref{LOB}) but with $p=q+1$, then the
suggested hypergeometric approximant is
\[
Q\left(  z\right)  \sim c_{0}\ _{\text{ }3}F_{2}(a_{1},a_{2},a_{3};b_{1}%
,b_{2};\sigma z).
\]
$\ c_{0}$ $_{\text{ }3}F_{2}(a_{1},a_{2},a_{3};b_{1},b_{2};\sigma z)$ has the expansion:%

\begin{equation}
\ c_{0\text{ }3}F_{2}(a_{1},a_{2},a_{3};b_{1},b_{2};\sigma z)=c_{0}+c_{0}%
\frac{a_{1}a_{2}a_{3}\sigma}{b_{1}b_{2}}z+c_{0}\frac{a_{1}\left(
1+a_{1}\right)  a_{2}\left(  1+a_{2}\right)  a_{3}\left(  1+a_{3}\right)
\sigma^{2}}{b_{1}\left(  1+b_{1}\right)  b_{2}\left(  1+b_{2}\right)  }%
z^{2}+O\left(  z^{3}\right)
\end{equation}
Matching this expansion with the series in Eq.(\ref{Qpert}), we get the
following set of equations:%
\begin{align*}
c_{0}\frac{a_{1}a_{2}a_{3}\sigma}{b_{1}b_{2}}  &  =c_{1},\\
\text{ }c_{0}\frac{a_{1}\left(  1+a_{1}\right)  a_{2}\left(  1+a_{2}\right)
a_{3}\left(  1+a_{3}\right)  \sigma^{2}}{b_{1}\left(  1+b_{1}\right)
b_{2}\left(  1+b_{2}\right)  }  &  =c_{2},
\end{align*}
also the the numerator and the denominator parameters are constrained by the
large order relation:%
\[
a_{1}+a_{2}+a_{3}-(b_{1}+b_{2})-1=b.
\]
This set of three coupled equations is to be solved for the unknown parameters
$b_{1},b_{2}$ and $\sigma$. Note that the parameter $\sigma$ can be obtained
from quasi-classical methods but it is out of the scope of this work.

For the model with the ground state perturbation series in Eq.(\ref{Jpert}),
we have $a_{1}=\frac{-3}{2}$, $a_{2}=\frac{-1}{4}$ and $a_{3}=1 $
\cite{zin-borel} while the large order parameter $b=-\frac{3}{2}$
\cite{Large-strong}. Thus the solution of the above set of equations for that
model yields the results $b_{1}=-0.60310956052580091716$, $b_{2}=$
$0.35310956052580091716$ and $\sigma=-0.560266190804551029423$. Accordingly we
have the second order approximant:%
\begin{equation}
E_{0}^{J}\simeq0.37 _{\text{ }3}F_{2}\left(  \frac{-3}{2},\frac{-1}%
{4},1;-0.60,0.35;-0.56J\right)  . \label{3F2}%
\end{equation}
One can involve more perturbative terms as input by going to $_{\text{ }%
4}F_{3}$,$_{\text{ }5}F_{4}$ $... $ and so on. To test the accuracy of the
algorithm, we compare its prediction with exact (numerical results) from
Ref.\cite{bendr,bendrII} in table \ref{Eg} . Note that the vacuum energy for the
Hamiltonian in Eq.(\ref{gx3}) and that in Eq.(\ref{Jx3}) are related as
$E_{0}^{g}=J^{\frac{-1}{4}}\left(  E_{0}^{J}- \frac{1}{3}{m^{6}}{J^{\frac{3}{2}}}\right)  $ and $J=\frac{im^{4}%
}{g^{\frac{4}{5}}}$ while the coupling $\lambda$ in Refs.\cite{bendr,bendrII} is
related to $g$ by the relation $g=288\lambda^{2}$. From table \ref{Eg}, one
can realize that the accuracy of the algorithm is improved form order to order.

\begin{table}[pth]
\caption{{\protect\scriptsize { Comparison of our prediction for $E_{0}%
^{g}$ and numerical results $E_{exact}$ from Ref.\cite{bendr,bendrII}. We get first the hypergeometric
approximations  $_{3}F_{2} , \ _{4}F_{3},\ _{5}F_{4}$ and $_{6}F_{5}$ for the
perturbation series of $E_{0}^{J}$ and then transform it to $E_{0}^{g}$. Note
that $J=\frac{im^{4} }{g^{\frac{4}{5}}}$ and $g=288\lambda^{2}$ while we set
$m=1$.}}}%
\label{Eg}%
\begin{tabular}
[c]{|l|l|l|l|l|l|}\hline
$\ \ \lambda$ & \ \ $_{\text{ }3}F_{2}$ & \multicolumn{1}{c|}{$_{\text{ }%
4}F_{3}$} & \ \ $_{\text{ }5}F_{4}$ & \ \ $_{\text{ }6}F_{5}$ & Exact\\\hline
0.015625 & 0.682387 & 0.504794 & 0.501965 & 0.502697 & 0.502621\\\hline
\ 0.03125 & \multicolumn{1}{c|}{0.534941} & \multicolumn{1}{c|}{0.510201} &
0.509934 & 0.509978 & 0.50998\\\hline
\ \ 0.0625 & 0.536264 & 0.533944 & 0.533931 & 0.533932 & 0.53393\\\hline
\ \ \ 0.125 & 0.595069 & 0.594916 & 0.594915 & 0.594915 & 0.59492\\\hline
\ \ \ \ 0.25 & 0.712944 & 0.712936 & 0.712936 & 0.712936 & 0.71294\\\hline
\ \ \ \ 0.5 & 0.900258 & 0.900258 & 0.900258 & 0.900258 & 0.90026\\\hline
\ \ \ \ \ 1 & 1.16745 & 1.16745 & 1.16745 & 1.16745 & 1.16746\\\hline
\ \ \ \ \ 2 & 1.53077 & 1.53077 & 1.53077 & 1.53077 & 1.53078\\\hline
\end{tabular}
\end{table}

One can obtain the edge critical exponent and the critical coupling of the
theory by noting that the hypergeometric functions $_{\text{ }p}F_{p-1}\left(
{a_{1},......a_{p};b_{1},........b_{q};\sigma z}\right)  $ have a power-law
behavior around the tip of the branch cut (starting from ${\sigma z=1}$ to  ${\sigma z\rightarrow\infty}$)  in the form
\cite{HTF,Math,Bos-Hub}:%
\begin{equation}
_{\text{ }p}F_{p-1}\left(  {a_{1},..a_{p};b_{1},..b_{p-1};\sigma z}\right)
-{}_pF_{p-1}\left(  {a_{1},..a_{p};b_{1},..b_{p-1};1}\right)
\propto(1-{\sigma z)}^{y},
\end{equation}
where
\begin{equation}
y=\sum_{i=1}^{p-1}b_{i}-\sum_{i=1}^{p}a_{i}=-(b+1). \label{exponent}%
\end{equation}
This means that as $J\rightarrow J_{c}=\frac{1}{\sigma}$ we have
\[
E_{0}^{J}(J)-E_{0}^{J_{c}}=\propto(1-{\sigma J)}^{\frac{1}{2}}.
\]
The critical coupling $J_{c}$ from the second order approximant in
Eq.(\ref{3F2}) is thus $J_{c}=\allowbreak-1.\,\allowbreak784\,9$ compared to
$J_{c}=-1.3510$ from ODM rsummation at the 150th order from
Ref.\cite{zin-borel}. In fact, at the fifth order approximant ($_{\text{ }%
6}F_{5}$) we obtained a precise value for the critical coupling $J_{c}$ as
shown in table \ref{JC}.

\begin{table}[pth]
\caption{{\protect\scriptsize { The hypergeometric $_{3}F_{2} , \ _{4}%
F_{3},\ _{5}F_{4}$ and $_{6}F_{5}$ predictions for the critical coupling
$J_{c}$ compared to the $150^{th}$ order of the ODM method in
Ref.\cite{zin-borel}. All approximants predict the same exact critical
exponents as shown because they depend solely on the large order parameter
$b=-3/2$. The fifth order approximant ($_{\text{ }6}F_{5}$ ) gives a very
precise critical coupling as shown in the table.}}}%
\label{JC}%
\begin{tabular}
[c]{|l|l|l|l|l|}\hline
Approximant & \ \ \ $J_{c}$ & \ $\nu_{c}$ & $\delta$ & \ $\gamma$\\\hline
\ \ \ \ \ $_{3}F_{2}$ & -1.78487 & 1/2 & -2 & 3/2\\\hline
\ \ \ \ \ $_{4}F_{3}$ & -1.30267 & 1/2 & -2 & 3/2\\\hline
\ \ \ \ \ $_{5}F_{4}$ & -1.32908 & 1/2 & -2 & 3/2\\\hline
\ \ \ \ \ $_{6}F_{5}$ & -1.35062 & 1/2 & -2 & 3/2\\\hline
\ \ \ \ \ ODM & -1.351 0 &\ \ - &  \ - &\ \  -\\\hline
\end{tabular}
\end{table}
What is really impressive is that according to our prediction, the critical
exponent $\nu_{c}=\frac{1}{2}$ which is extracted from the relation $\left(
E_{0}^{J}(J)-E_{0}^{J_{c}}\right)  \propto(1-{\sigma J)}^{d\nu_{c}}%
$\cite{Kaku}. This result is exact \cite{LYsing1} $\left(  d=1\right)  $ and
does not depend on the order of approximation but on the other hand depends
solely on the large order parameter $b=-\frac{3}{2}$. This is clear from
Eq.(\ref{large-b}) where we find $y=-1-(-\frac{3}{2})=\frac{1}{2}$. It has been
shown in Ref.\cite{LYsing1} that the edge critical exponent $\nu_{c}$ for one
dimensional Ising model is $\nu_{c}=\frac{1}{2}$ exactly the same value we
obtained. \ Note that \ here we used the
scaling relation $\left(  E_{0}^{J}(J)-E_{0}^{J_{c}}\right)  \propto
\zeta_{gap}^{-d}\propto (J-J_c)^{-d\nu_{c}}$ \cite{Kaku} , where $\zeta_{gap}^{-d}$
is the correlation length. Up to the best of our knowledge this is the first
time to get exact critical exponents from only the knowledge of the
large-order parameters. Note that the square root singularity of the ground
state energy near the critical coupling $J_{c}$ has been suggested based on
the analysis of the calculations in Ref.\cite{zin-borel} but here we get it exactly.

The ground state energy or equivalently the effective potential is well known
to be the generating functional of the one-particle irreducible amplitudes
\cite{Peskin}. Accordingly, one can obtain other amplitudes like magnetization
(vacuum expectation value) and magnetic susceptibility   for
instance from successive differentiation with respect to $\left(  \frac{1}%
{2}iJ\right)$. Thus the vacuum condensate $v$ is given by
\[
v=\frac{\partial E_{0}^{J}}{\partial\left(  \frac{1}{2}iJ\right)  },
\]
where the hypergeometric approximant for $E_{0}^{J}$ is given by
\[
E_{0}^{J}\approx c_{0}\text{ \ }_{\text{ }p}F_{p-1}\left(  {a_{1}%
,......a_{p};b_{1},........b_{q};\sigma J}\right)
\]
Note also that%

\[
\frac{\partial}{\partial z} \text{} _{p}F_{q}\left(  a_{1},........a_{p}%
;b_{1}......,b_{q};z\right)  =\frac{\prod_{j=1}^{p}a_{j}}{\prod_{j=1}^{q}%
b_{j}}\,_{p}F_{q}\left(  a_{1}+1,......a_{p}+1;b_{1}+1,....b_{q}+1;z\right)
.
\]
We found that the vacuum expectation value is negative imaginary as it is well
known for such $\mathcal{PT}$-symmetric model \cite{ivev}. The derivative of
the hypergeometric function is thus another hypergeometric function but with
every numerator and denominator parameter is increased by $1$. The exponent of
the power-law behavior of the derivative will thus decrease by $1$. Thus near
the critical point, the vacuum condensate $v$ has a power-law behavior of the
from:
\[
v\left(  J\right)  -v(J_{c})\propto\left(  J-J_{c}\right)  ^{\frac{1}{\delta}%
},
\]
where $\frac{1}{\delta}=y-1$ and $y$ is defined in Eq.(\ref{exponent}). Since
we obtained $y=\frac{1}{2}$ for the model under consideration, we get
$\delta=-2$. This is again the exact exponent reported in Ref. \cite{LYsing1}.

The magnetic susceptibility $\chi$  is given
by:%
\[
\chi=\frac{\partial^{2}E_{0}^{J}}{\partial\left(  \frac{1}{2}iJ\right)  ^{2}%
}.
\]
Accordingly, $\chi$ has the power-law behavior%

\[
\chi\propto\left(  J-J_{c}\right)  ^{-\gamma},\text{ \ }-\gamma=y-2=\frac
{-3}{2}%
\]
This result is in accordance with scaling relations that relate $\gamma$ to
$\delta$ as \cite{Pal}
\[
-\gamma=\frac{1-\delta}{\delta}=\frac{-3}{2}.
\]
So again  our result is exact. Since all critical exponents here are obtained from the large-order parameter $b$, the results suggest the universality of such parameter.
\section{Hypergeometric Approximation for the  HT-expansion of the  susceptibility  within the SC LATTICE}
\label{HT-SC}
In the previous section, we considered a $0+1$ dimensional quantum field example for which the parameter $b$ is known exactly and showed that the critical exponent is solely determined by $b$. To test the validity of our conjecture for a  higher dimensional example, one should first note that the mathematical structure of the strong-coupling expansion of   the $\phi^4$ lattice field theory is equivalent to the HT-expansion of the Ising model \cite{Yamada2007}. With no loss of generality, for a field theory in higher dimensions, one can consider  the high-temperature expansion within the Ising model to test     the universality of the large-order parameter $b$. Up to the best of our knowledge, the large-order asymptotic behavior is not known exactly for such expansion. However, we choose such a case because  the expansion is known up to high order and thus an accurate prediction of $b$ can be obtained using the hypergeometric approximants. Since the high temperature   expansion within the Ising model  has a finite radius of convergence, the expanded
quantity can be approximated by the hypergeometric approximant $_{\text{ }%
p}F_{p-1}\left(  {a_{1},......a_{p};b_{1},........b_{q};\sigma z}\right)  $
which has the large-order asymptotic behavior of the form $\sigma^{n}%
n^{b}\left(  1+O\left(  \frac{1}{n}\right)  \right)  $. A very good
approximation  of the parameters $b$ and $\sigma$ can be extracted from the
available many terms of the perturbation series. The HT-expansion up to
$O(\beta^{25})$ for the susceptibility of the $SC$ lattice ($S=\frac{1}{2})$
is given by \cite{HT1}:%

\begin{align}\label{HT-SCs}
\chi\left(  \beta\right)   &  =\sum_{i=0}^{\infty}c_{i}\beta^{i}%
=1+6\beta+30\beta^{2}+148\beta{{}^3}+706\beta^{4}+\dots\dots\dots\nonumber\\ 
&  +\frac{93592219478518291774477772}{3093594879375}\beta^{20}\nonumber\\
&  +\frac{8972803527064109944099241768}{/64965492466875}\beta{{}^2} {{}^1}\nonumber\\
&  +\frac{5296430224856866468505272024}{8407299025125}\beta{{}^2}{{}^2}\nonumber\\
&  +\frac{47221618622049399307213422740992}{16436269594119375}\beta{{}^2}{{}^3}\nonumber\\
&  +\frac{37975296352037116774213386661036}{2900518163668125}\beta^{24}\nonumber\\
&  +\frac{73538934029908819825899053186296808}{1232720219558953125}\beta
^{25}+\dots\dots\dots
\end{align}
One can go and parametrizes the approximant $_{\text{ }p}F_{p-1}\left(
{a_{1},......a_{p};b_{1},........b_{q};\sigma z}\right)  $ starting from the
lowest orders and watch the convergence as more orders are employed. However,
according to our conjecture the critical indices are encoded in the large
orders. Accordingly, to obtain fast accurate predictions for the large-order
parameters $b$ and $\sigma$, it would be better to parametrize the the
approximant $_{\text{ }p}F_{p-1}\left(  {a_{1},......a_{p};b_{1}%
,........b_{q};\sigma z}\right)  $ starting from highest to lower orders.
Also, as explained in Ref.\cite{HT2}, the oscillation in the sign of the term
$O\left(  \frac{1}{n}\right)  $ in the large-order form demands us to treat
odd (or even) terms separately. Accordingly, the parametrization of the
approximant can go through matching the relations (for odd terms)
\begin{align*}
R_{n}^{odd}  &  =\frac{c_{i}}{c_{i-2}},\text{ }i=25,23,21,......\\
R_{25}  &  =\frac{c_{25}}{c_{23}},R_{23}=\frac{c_{23}}{c_{21}},R_{21}%
=\frac{c_{21}}{c_{19}},....
\end{align*}
For instance, the four parameters approximant $\chi\simeq {}_{2}F_{1}\left(
{a_{1},a_{2};b_{1};\sigma z}\right)  $ can be parametrized as:%

\begin{align}
\frac{(a_{1}+23)(a_{1}+24)(a_{2}+23)(a_{2}+24)}{(b_{1}+23)(b_{1}+24)}%
\sigma^{2}  & =\frac{9192366753738602478237381648287101}%
{737837790969521864175209730328}\nonumber\\
\ \frac{(a_{1}+21)(a_{1}+22)(a_{2}+21)(a_{2}+22)}{(b_{1}+21)(b_{1}+22)}%
\sigma^{2}  & =\frac{11805404655512349826803355685248}%
{1121600440883013743012405221}\nonumber\\
\frac{(a_{1}+19)(a_{1}+20)(a_{2}+19)(a_{2}+20)}{(b_{1}+19)(b_{1}+20)}%
\sigma^{2}  & =\frac{4486401763532054972049620884}{512422018420605679707007}%
\label{large-param}\\
\frac{(a_{1}+17)(a_{1}+18)(a_{2}+17)(a_{2}+18)}{(b_{1}+17)(b_{1}+18)}%
\sigma^{2}  & =\frac{2049688073682422718828028}{286754736324335778241}%
\nonumber
\end{align}
Solving these equations for the parameters $a_{1},a_{2},b_{1}$ and $\sigma$
gives the result:%
\begin{align*}
\sigma & =4.5108,\text{ \ }a_{1}=-23.2765-1.20827\times10^{-8}i\\
a_{2}  & =1.24944,\text{ \ }b_{1}=-23.2765-1.20826\times10^{-8}i.
\end{align*}
The large-order parameter $b$ is related to the numerator parameters $a_{i}$,
and the denominator parameter $b_{i}$ as:%
\begin{align*}
b\bigskip & =\sum_{i=1}^{p}a_{i}-\sum_{i=1}^{p-1}b_{i}-1\\
& =0.\,\allowbreak249\,4-1.0\times10^{-13}i\allowbreak\\
& \simeq0.\,\allowbreak249\,4.
\end{align*}
Thus the $\gamma$ exponent is given by $\gamma=b+1=\allowbreak1.249\,4$
compared to $1.237$ in Ref.\cite{HT2}. Also, the critical inverse temperature
$\beta_{c}=\frac{1}{\sigma}=$  $0.221\,61$ compared to $0.22165$ from
Ref.\cite{HT2}. So again the large order parameter $b$ determines the critical
exponent. Note that, very close results are obtained by treating the even terms
but we prefer the odd ones as the last term is odd and it is assumed to be
more effective in determining the large-order parameter $b$.
The six parameters approximant $\chi\simeq{}_{3}F_{2}\left(  {a_{1},a_{2}%
,a_{3};b_{1},b_{2};\sigma z}\right)  $ can also be parametrized through the
following relations:%

\begin{align}
\frac{(a_{1}+23)(a_{1}+24)(a_{2}+23)(a_{2}+24)(a_{3}+23)(a_{3}+24)}%
{(b_{1}+23)(b_{1}+24)(b_{2}+23)(b_{2}+24)}\sigma^{2}  & =20.7642\nonumber\\
\ \frac{(a_{1}+21)(a_{1}+22)(a_{2}+21)(a_{2}+22)(a_{3}+21)(a_{3}+22)}%
{(b_{1}+21)(b_{1}+22)(b_{2}+21)(b_{2}+22)}\sigma^{2}  & =20.801382\nonumber\\
\frac{(a_{1}+19)(a_{1}+20)(a_{2}+19)(a_{2}+20)(a_{3}+19)(a_{3}+20)}%
{(b_{1}+19)(b_{1}+20)(b_{2}+19)(b_{2}+20)}\sigma^{2}  & =20.845921\nonumber\\
\frac{(a_{1}+17)(a_{1}+18)(a_{2}+17)(a_{2}+18)(a_{3}+17)(a_{3}+18)}%
{(b_{1}+17)(b_{1}+18)(b_{2}+17)(b_{2}+18)}\sigma^{2}  & =20.900230\nonumber\\
\frac{(a_{1}+15)(a_{1}+16)(a_{2}+15)(a_{2}+16)(a_{3}+15)(a_{3}+16)}%
{(b_{1}+15)(b_{1}+16)(b_{2}+15)(b_{2}+16)}\sigma^{2}  & =20.967904\\
\frac{(a_{1}+13)(a_{1}+14)(a_{2}+13)(a_{2}+14)(a_{3}+13)(a_{3}+14)}%
{(b_{1}+13)(b_{1}+14)(b_{2}+13)(b_{2}+14)}\sigma^{2}  & =21.054549.\nonumber
\end{align}
The solution of this set of equations gives the values:  
\begin{align*}
\sigma & =4.51093,\text{ \ }a_{1}=-15.4176-2.76717\times10^{-9}i\\
a_{2}  & =1.24892,\text{ }a_{3}=-17.7923-2.96486\times10^{-9}i\\
\text{\ }b_{1}  & =-15.4176-2.76719\times10^{-9}i,\text{ }\\
b_{2}  & =\text{\ }-17.7923-2.9648\times10^{-9}i.
\end{align*}
This gives $b=0.2489-3.76237\times10^{-14}i$ and $\beta_{c}=\allowbreak
0.221\,68$. So again the large order parameter $b$ determines the critical
exponent. Also, for the eight parameters hypergeometric approximants $_{4}F_{3}\left(
{a_{1},a_{2},a_{3},a_{4};b_{1},b_{2},b_{3};\sigma\beta}\right)  ,$ we obtained
the following parameters values:
\begin{align*}
\sigma & =4.51101,\text{ \ }a_{1}=-15.8686-5.10971\ \times10^{--7}i\\
a_{2}  & =1.2483819+2.37927\times10^{-10}i\ ,\text{ }a_{3}%
=-11.3347-4.45387\times10^{-8}i\\
a_{4}  & =-13.6312-2.11442\times10^{-7}i,\text{\ }b_{1}=-11.3345-4.46736\times
10^{-8}i,\text{ }\\
b_{2}  & =\text{\ }-13.6314-2.11099\times10^{-7}i,b_{3}=\text{\ }%
-15.8686-5.11116\times10^{-7}i,
\end{align*}
 This leads to the result $b=0.24838+1.74976\times10\symbol{94}%
^{-10}i$ and $\beta_{c}=\frac{1}{4.51101}=\allowbreak0.221\,68$. In table \ref{bSC}, we listed the  hypergeometric predictions  for $b,\gamma$ and $\beta_c$.
\begin{table}[H]
\caption{{\protect\scriptsize { The hypergeometric approximants $_{2}F_{1} , \ _{3}%
F_{2}$ and $\ _{4}F_{3}$  predictions for the large-order parameters $b$ and $\beta_c=\frac{1}{\sigma}$ of the HT-series expansion of the susceptibility of the SC lattice. These approximants are parametrized using the last odd orders of the series. Also, the critical exponent $\gamma=b+1$ is listed  where it can  be compared  
 to $\gamma=1.237$ in Ref.\cite{HT2} while the critical inverse temperature $\beta_{c}$   can be compared with  $\beta_c=0.22165$ from the same reference.}}}%
\label{bSC}%
\begin{center}

\begin{tabular}{|l|l|l|l|}
\hline
Approximant & \ \ \ $b$ & \ \ \ $\gamma$ &\ \ \ $\beta_c$ \\ \hline
\ \ \ ${}_2F_1$ & 0.2494 & 1.2494 & 0.22169 \\ \hline
\ \ \ ${}_3F_2$ & 0.2489 & 1.2489 & 0.22168 \\ \hline
\ \ \ ${}_4F_3$ & 0.24838 & 1.24838 & 0.22168 \\ \hline
\end{tabular}
\end{center}
\end{table}
  Note that the series in Eq.(\ref{HT-SCs}) is not alternating in sign and thus the hypergeometric approximant is suffering from a problem like non-Borel summability. In fact the hypergeometric approximant has a branch cut starts from $\beta=1/\sigma$ to infinity. This means that it can (accurately) describe the high temperature phase only. For the low temperature phase, the hypergeometric approximants suffer from the existence of Stokes phenomena which can be cured \cite{Stokes2} but it is out of the scope of this work.  
\section{Hypergeometric Approximation for the HT-expansion of  the susceptibility of the SQ Ising model}\label{HT-SQ}

The HT-expansion for the susceptibility of the spin-half square lattice is given
by \cite{HT1}: 

\begin{align}
\chi\left(  \beta\right)   &  =1+4\beta+12\beta^{2}+\frac{104}{3}\beta{{}^3}
+92\beta^{4}+\dots\dots\dots\nonumber\\
&  +\frac{415782048556042942544}{3093594879375}\beta^{20}+\frac
{4735391065845611373232}{14992036723125}\beta{{}^2}{{}^1}\nonumber\\
&  +\frac{529562920319138348552816}{714620417135625}\beta{{}^2}{{}^2}
+\frac{85616154520095267692857616}{49308808782358125}\beta{{}^2}{{}^3}\nonumber\\
&  +\frac{66773068948180944546678128}{16436269594119375\ }\beta^{24} 
+\frac{3192145249472459217984684656}{336196423516078125\ }\beta^{25}+\dots\dots\dots,
\end{align}
  As long as we are interested only in  large-order parameters, one can fit
the ratio $R_{n}=\frac{C_{n}}{C_{n-1}}$   of the above series with the corresponding ratio of expansion coefficients of the hypergeometric approximant but for large orders. In fact, if this series to have a finite radius of convergence, then for large $n$ it has to fit the corresponding
ratio from the expansion of the hypergeometric approximant as:
\begin{align*}
R_{n}  &  \simeq\frac{(\sigma)^{n}n^{b}}{(\sigma)^{n-1}\left(  n-1\right)
^{b}}\\
&  =\sigma\frac{\ \ 1}{\ \left(  1-\frac{1}{n}\right)  ^{b}}\\
&  \simeq\sigma\left(  1+b\left(  \frac{1}{n}\right)  \right)  .
\end{align*}
\begin{figure}[htp]
\begin{center}
\epsfig{file=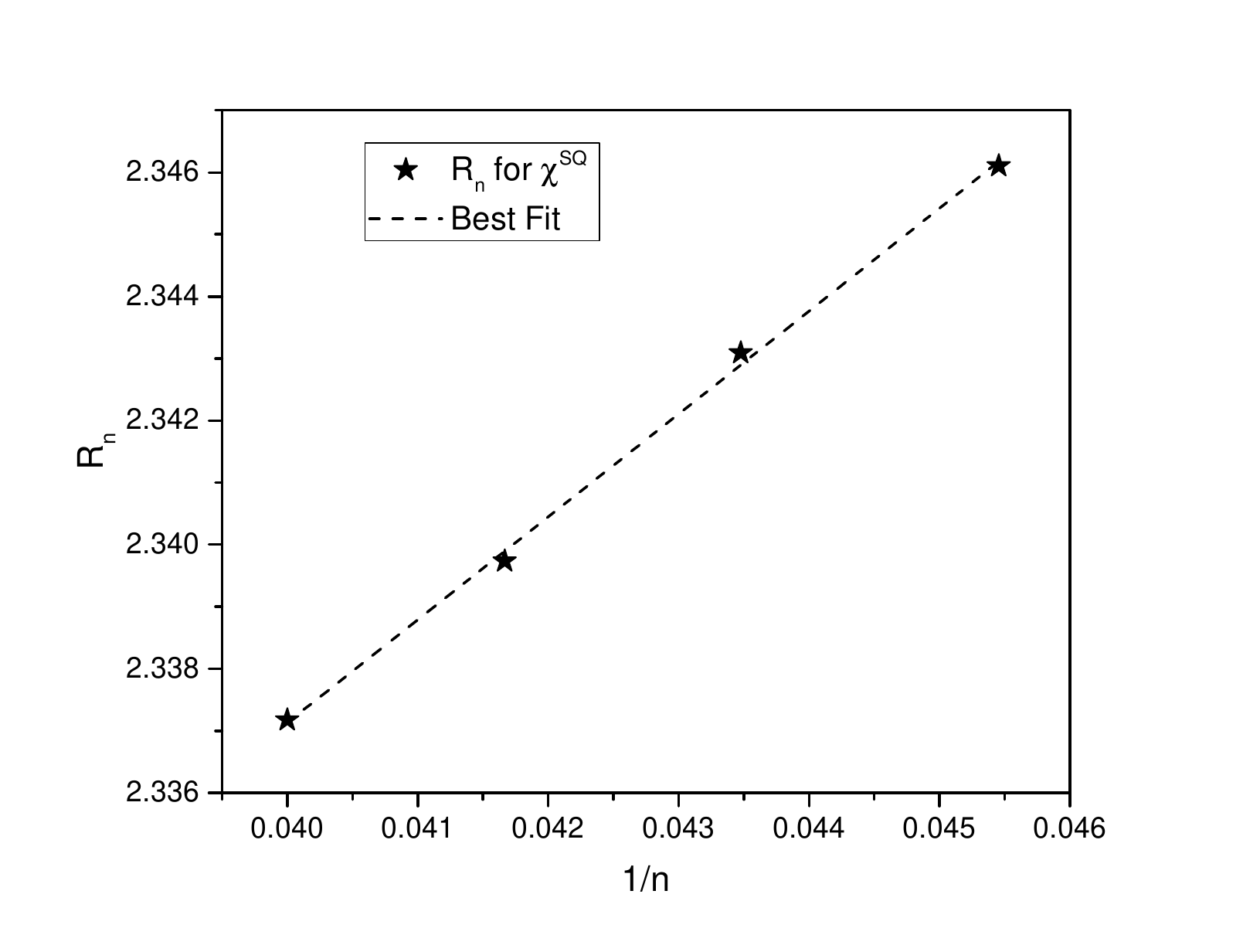,width=0.65\textwidth}
\end{center}
\caption{\textit{In this figure, we plot the ratio $R_n$ for the  HT series expansion of the susceptibility  of the spin-half square lattice at large orders. The data has a straight line fit of the form $R_n=1.7025(1/n) + 2.269$ which predicts the values $\sigma=2.269$ and $b=0.75033$.}} 
\label{xi-SQ} 
\end{figure}
In Fig.\ref{xi-SQ}, we plot the ratio $R_{n}$ versus $\frac{1}{n}$ from which we
obtained the large order parameters values $b=0.750\,33$ and $\sigma=2.269$. These values lead to the critical exponent
$\gamma=b+1=1.750\,33$ and critical $\beta$ as $\beta_{c}=\frac{1}{\sigma
}=0.440\,72.$ The exact values are well known to be $\gamma=\nu\left(
2-\eta\right)  =1.75$ \cite{exactexp} , $\ \beta_{c}=0.4407$ \cite{Exactbeta}. Again, this example assures the universality of the large-order parameter $b$ for the HT series expansion.

\section{Summary and Conclusions}\label{conc}

Near the tip of the branch cut, the hypergeometric functions   ${}_{k+1}F_k$ have a power-law behavior   similar to a physical quantity   (magnetization for instance)   near second-order phase transition where it follows the form $Q(\beta)\propto(1-\frac{\beta}{\beta_c})^\psi$. The exponent $\psi$ has been proven to follow the relation $\psi=-(b+1)$, where $b$ is the large-order parameter of the series. This link between $b$ and $\psi$ is in complete agreement with the theorem of Darboux.  Accordingly, one can determine the exact critical exponent provided that we know the   large-order parameter $b$ of a series expansion with finite radius of convergence. For the weak coupling expansion which is generated around an essential singularity,  Darboux theorem is not applicable and thus a direct link between $\psi$ and $b$ is missing. On the other hand, the strong-coupling (HT) expansion is well known to have a finite radius of convergence. Accordingly, one can expect that the large-order parameter $b$ of  the strong-coupling (HT) expansion is universal. In fact there are techniques in quantum field theory that might enable us  to get exact values for the parameter $b$ and in this case exact critical exponents can be extracted. 

We tested our conjecture  about universality of $b$ using the Yang-Lee model in 0+1 space-time dimensions as well as the HT-series expansion for Ising model. The strong coupling expansion of the Bose-Hubbard model is agreeing with our thoughts too \cite{Bos-Hub,Bos-Hub1}.

 We  started from the large-order behavior of the perturbation
series of the ground state energy of the Yang-Lee model to select the suitable
hypergeometric approximant. We realized that the large-order behavior does not
have a growth factor $n!$ and found that the hypergeometric functions
$\text{{}}_{p}F_{p-1}$ do have the same form of the given large-order
behavior. This recommends them to be the most suitable hypergeometric
approximants for the given series. We set a constraint on the numerator and
denominator parameters based on matching both large-order behaviors. The
large-order constraint on the parameters has accelerated the convergence which
has been tested by calculating up to fifth order approximants ($\text{{}}%
_{6}F_{5}$) and found that they yield very precise predictions of the ground
state energy compared to exact (numerical) results from literature.

$\mathcal{PT}$- symmetry breaking has been investigated by noting that the
hypergeometric functions $\text{}_{p}F_{p-1} \left(  a_{1},a_{2},....a_{p};
b_{1},b-2,...b_{p-1};-\sigma z\right)  $ have branch cut starting at $-\sigma z=1$ and extends to $-\sigma z=\infty$. Near the branch point $-\sigma z=1$ , the
hypergeometric function has a power-law behavior from which we were able to
get the exact $\nu_{c}$ critical exponent and a very precise value for the
critical coupling. We found that the exact
critical exponent can be extracted from the large-order parameter $b$ which up to
the best of our knowledge is the first time to extract exact critical
exponents directly from large-order parameters. This prediction might open the
door to directly extract exact critical exponents from obtaining the
large-order behavior of a given perturbation series (strong-coupling (HT)).

Since the ground state energy serves as the effective potential, it enables us
to obtain all the critical exponents (exact) by successive differentiation of
the ground state energy (effective potential) with respect to the external
magnetic field.

The critical exponents of the Yang-Lee model are always stressed within the
investigation of the zeros of the partition function which very recently has
been exposed to experimental investigations \cite{yang-lee-exp}. In this work
we extracted the edge exponent $\delta$ from the dependance of the order
parameter on the external magnetic field. In Ref.\cite{abo-yang}, we have
shown that the critical point of $\mathcal{PT}$- symmetry breaking is in fact
a Yang-Lee edge singularity. So our results here might lead to experimental
investigation of Yang-Lee edge singularity via testing $\mathcal{PT}$-symmetry
breaking as well as watching the behavior of the order parameter near the
point of symmetry breaking.

Our conjecture has been tested also by considering the HT-series expansion of the SC and SQ lattices of the spin half Ising model. From the many orders known for susceptibility expansion, we were able to get accurate values for the large-order parameter $b$. The universality of $b$ has been also assured for these examples by comparing the extracted critical exponent with the  well known one of the Ising model. We also obtained a very accurate critical temperature from the approximate  large-order behavior of the given series.

 The main message of this work is to draw the attention to the importance of determining  the large-order asymptotic behavior of the strong-coupling (HT) expansion in field theory. There exists well known methods to do that for a series expansion and thus if we were able to get exact values for $b$, it will lead to the first determination of an exact critical exponent in three dimensions.  For the $O(2)$ symmetric $\phi^4$ theory this will resolve  the current $\lambda $-point dispute between theory and experiment \cite{dispute,dispute1,abodispute}.
 
\bibliography{UniversalityR1}
\end{document}